# Uncovering shared common genetic risk factors for various aspects of complex disorders captured in multiple traits

**Summer S. Han**

Department of Statistics, Yale University,

24 Hillhouse Ave., New Haven, CT 06520, USA

email: seongmin.han@yale.edu

**Elena L. Grigorenko**

Child Study Center, Yale School of Medicine, Yale University

230 South Frontage Road, New Haven, CT06519-1124

email: elena.grigorenko@yale.edu

and

**Joseph T. Chang**

Department of Statistics, Yale University,

24 Hillhouse Ave., New Haven, CT 06520, USA

email: joseph.chang@yale.edu

# Summary

Identifying shared genetic risk factors for multiple measured traits has been of great interest in studying complex disorders. Marlow's (2003) method for detecting shared gene effects on complex traits has been highly influential in the literature of neurodevelopmental disorders as well as other disorders including obesity and asthma. Although its method has been widely applied and has been recommended as potentially powerful, the validity and power of this method have not been examined either theoretically or by simulation. This paper establishes the validity and quantifies and explains the power of the method. We show the method has correct type 1 error rates regardless of the number of traits in the model, and confirm power increases compared to standard univariate methods across different genetic models. We discover the main source of these power gains is correlations among traits induced by a common major gene effect component. We compare the use of the complete pleiotropy model, as assumed by Marlow, to the use of a more general model allowing additional correlation parameters, and find that even when the true model includes those parameters, the complete pleiotropy model is more powerful as long as traits are moderately correlated by a major gene component. We implement this method and a power calculator in software that can assist in designing studies by using pilot data to calculate required sample sizes and choose traits for further linkage studies. We apply the software to data on reading disability in the Russian language.

# 1. Introduction

Identifying shared genetic risk factors on various aspects of complex disorders has been of great interest to researchers studying neurodevelopmental, psychiatric and cognitive disorders as well as other types of disorders such as asthma and obesity (Bates, 2006; Fisher, 2006; Gayan and Olson, 2001; Gayan et al., 2005; Hansell et al., 2005; Hohnen and Stevenson, 1999; Loo et al., 2004; Stevenson et al., 1993; Zhao et al., 2008).

In studying such disorders using multiple phenotypes, several multivariate methods have been proposed, and have been found to be more powerful than a standard univariate method, because they exploit extra information related to correlations among multiple traits (Almasy, Dyer and Blangero, 1997; Amos, de Andrade and Zhu, 2001; Amos and de Andrade, 2001; Boomsma and Dolan, 1998; Evans, 2002; Evans et al., 2004; Schmitz, Cherny and Fulker, 1998; Vogler et al., 1997; Williams et al., 1999a; Williams et al., 1999b). Although these multivariate approaches are different in methodological details, the main idea underlying these methods is to test for QTL influence on multiple phenotypes simultaneously through a joint null hypothesis that asserts that the QTL has no effect on any of the traits. Rejecting this joint null hypothesis suggests that the locus under consideration is a risk factor for at least one of the traits included in the multivariate model but does not suggest which specific traits might be influenced by the QTL. Thus, clarifying shared genetic effects among traits is beyond the scope of these methods.

Marlow et al. (2003) took a different multivariate approach for discovering shared genetic effects on diverse cognitive skills related to developmental dyslexia. Their method (hereafter referred to as MTST, for multivariate trait-specific test) tests the QTL effect on each individual trait through a set of marginal null hypotheses, still fully exploiting correlation information among traits using a multivariate model. Table 1 displays the results of the MTST applied to six traits on the locus at 18p11.2 in their data, with the corresponding results of univariate tests (UT hereafter). While the UT results suggest that the QTL influence is specific only to the Reading trait (only the Reading trait is significant, at P=0.000009), the MTST results suggest the QTL is a generalist gene influencing multiple cognitive skills related to developmental dyslexia, with all traits displaying significance much increased over the UT results(P=0.000004~0.006).

Table 1. P values from Marlow et al. (2003) using the univariate test (UT) and the multivariate shared gene test (MTST) on a locus in Chromosome 18

| Trait | UT | MTST |
|---|---|---|
| Reading | 0.000009 | 0.000004 |
| Orthography-irregular | 0.02 | 0.000007 |
| Phonological decoding | 0.1 | 0.00006 |
| Orthography-choice | 0.12 | 0.00008 |
| Spelling skills | 0.49 | 0.0001 |
| Phonological awareness | 0.5 | 0.006 |

Marlow's method for detecting shared gene effects has been highly influential in the literature of neurodevelopmental, cognitive, and psychiatric disorders as well as other types of disorders that use multiple complex traits, and has been applied to several other studies (Bates, 2006; Cherny, 2008; SLI Consortium, 2004; Fisher, 2006; Hansell et al., 2005; Loo et al., 2004; Monaco, 2007; Schulte-Korne, 2002; Wright et al., 2008; Zhao et al., 2008). Although this method has had an impact on numerous studies, and has been recommended to be a potentially powerful method (Cherny, 2008; Fisher, 2006), its validity and power have not been examined either theoretically or by simulation.

In this paper, we study the validity and the power of the MTST by estimating type 1 error rates through simulations and by calculating asymptotic powers of the test using noncentrality parameters of chi-square distributions. We also provide explanations on the source of power gains achieved by the MTST over the UT. Our study shows the MTST has correct type 1 error rates, with asymptotic null distribution being chi-square with 1 degree of freedom regardless of the number of traits included in the multivariate model. Our noncentrality parameter calculations confirm power increases in most cases using the MTST method across different genetic models, selected to be well spread out over the parameter space. We discover the main source of these power gains is correlations among traits induced by a common major gene effect component. We compare the use of the complete pleiotropy model, as assumed by Marlow, to the use of a more general model allowing additional correlation parameters. We find that even when the true model includes those additional parameters, the complete pleiotropy model is more powerful as long as traits are moderately correlated by a major gene component, due to the fact that the more general model has increased degrees of freedom due to estimating more parameters.

We implement the MTST and a power calculator in software that can assist in designing studies by using pilot data to calculate required sample sizes and choose traits for further studies. We apply this test and power calculator to new data from the Yale Child Study Center on five traits related to reading disability in the Russian language. We identify a locus on 3q29, located near regions reported to influence written and spoken language disorders in other languages including Finnish (Nopola-Hemmi et al., 2001) and English(Stein et al., 2004). Our results also suggest the possibility of 3q29 being a shared common genetic risk factor for all five traits related to reading disability. We illustrate how our power calculator applies to this pilot data by calculating required sample sizes and choosing traits for designing a follow-up study, which reduced the required sample size by a factor of seven, compared to a standard power calculation based on univariate tests.

## 2. Background

### *Variance components methods*

### A univariate test (UT)

In a univariate model, the response variable includes one measured trait for each individual, modeled as a sum of three independent normally distributed random effects: an additive major gene effect, which is to be tested on each genetic marker, a polygenic effect, and an environmental effect with corresponding variance parameters $a^2$, $g^2$ and $e^2$, respectively. For simplicity, here let us assume one sibpair in each family. The trait vector for the $i$ th family follows a multivariate normal distribution: $y_i = (y_{i1}, y_{i2}) \sim N(\mu, \Sigma_i)$ with

$$\Sigma_i = \begin{bmatrix} a^2 + g^2 + e^2 & \pi_{i,12}a^2 + 2\phi_{i,12}g^2 \\ \pi_{i,12}a^2 + 2\phi_{i,12}g^2 & a^2 + g^2 + e^2 \end{bmatrix}$$

, where $\pi_{i,12}$ is the identity by descent (IBD) sharing proportion and $\phi_{i,12}$ is the kinship coefficient between individuals 1 and 2. We test the null hypothesis $H_0: a^2 = 0$ against the alternative $H_1: a^2 > 0$. The likelihood-ratio test statistic (LRT hereafter) is twice the difference between the log-likelihood of the full model and the model restricted according to the null hypothesis. Due to nonstandard boundary conditions (Self and Liang, 1987), the asymptotic null distribution of LRT is $\frac{1}{2}\chi_0^2 + \frac{1}{2}\chi_1^2$, that is, the mixture distribution of 50% point mass at zero (equivalent to $\chi_0^2$) and 50 % $\chi_1^2$.

## A multivariate shared gene test (MTST)

Multivariate variance components models are a natural extension of the above single-trait model, where the response variables include $k$ measured traits for each individual. Continuing to illustrate with the case of one sib-pair per family,
$y_i = (y_{i11}, y_{i12}, \ldots, y_{i1k}; y_{i21}, y_{i22}, \ldots, y_{i2k}) \sim N(\mu, \Sigma_i)$, where $y_{ijt}$ is the value of trait $t$ measured on individual $j$ in family $i$, and $\Sigma_i = \begin{bmatrix} A+G+E & \pi_{i,12}A + 2\phi_{i,12}G \\ \pi_{i,12}A + 2\phi_{i,12}G & A+G+E \end{bmatrix}$.

Here $G = (g_{ij})$ and $E = (e_{ij})$ are general $k \times k$ covariance matrices for the polygenic effect and environmental effect, respectively. For later notational convenience, we define the correlation coefficients $\rho_{g_{ij}} = \frac{g_{ij}}{\sqrt{g_{ii}g_{jj}}}$ and $\rho_{e_{ij}} = \frac{e_{ij}}{\sqrt{e_{ii}e_{jj}}}$ for $i \neq j$. The covariance matrix for the additive major gene effect, $A = (a_{ij})$, is assumed to be an outer product of the form
$A = \begin{pmatrix} a_1 & a_2 & \cdots & a_k \end{pmatrix}^T \begin{pmatrix} a_1 & a_2 & \cdots & a_k \end{pmatrix}$, so that the covariances take the form of $a_{ij} = a_i a_j$.

This model, also known as the *single factor model* (Evans et al., 2004; Vogler et al., 1997) or the *complete pleiotropic model* (Almasy et al., 1997; Amos et al., 2001; Amos and de Andrade, 2001; Kraft et al., 2003; Williams et al., 1999a; Williams et al., 1999b), arises when the dominance components of variance for the effects of a single major gene on each trait are assumed to be 0. Additive gene effects have been largely supported by twin and family data for several complex disorders including cognitive and psychiatric disorders such as ADHD and reading disability (Monaco, 2007; Schulte-Korne, 2002; Smalley, 1997) and this complete pleiotropic model has been extensively used (Amos et al., 2001; Amos and de Andrade, 2001; Evans, 2002; Evans and Duffy, 2004; Evans et al., 2004; Marlow et al., 2003; Monaco, 2007; Vogler et al., 1997). This model is the main subject of interest in this paper, and the term the MTST will refer to a test based on this model unless mentioned otherwise. Later, we will use the term "general model" for models without the complete pleiotropic constraint.

The MTST has a set of the null hypotheses, $H_{0i}: a_i = 0$ for $i = 1,\ldots,k$, for marginally testing if the given locus influences the $i$th trait as a major gene, using the LRT. For the asymptotic null distribution, the chi-square distribution with one degree of freedom has been used (e.g. Marlow et al. 2003; Monaco 2007).

## 3. Type 1 error estimation

We simulated the null distributions of the MTST in *k*-trait multivariate models for $k = 2,3,4,5,6$, with 2000 replications each. The simulation method is the same as the one used by Han and Chang (2008) for another type of multivariate test that jointly tests for major gene effects on all traits included in the model using the null hypothesis $H_0: a_1 = \cdots = a_k = 0$. In this paper, the null hypothesis for the MTST will be $H_0: a_1 = 0$, testing the major gene effect on the first trait, in the presence of nonzero major gene effects on the other traits, which in our simulations account for 20% of the variation of each trait other than the first. We conducted LRT using the Mx software (Neale et al., 1999) for each simulated dataset, and obtained the null distribution of the MTST by collecting the LRT values from all datasets. We then estimated Type 1 error rates for significance levels $\alpha = 0.01$ and 0.05—which have corresponding critical values 6.635 and 3.841 for $\chi_1^2$—by counting the fractions of datasets that give LRT values larger than the given critical values.

The estimated Type 1 error rates for $\alpha = 0.05$ for $k = 2, 3, 4, 5$, and 6 were 0.048, 0.053, 0.052, 0.053, and 0.049, respectively, and for $\alpha = 0.01$, they were 0.009, 0.014, 0.012, 0.012 and 0.011, suggesting the tests are valid. We also conducted Kolmogorov-Smirnov tests to compare the simulated null distributions of the MTST to $\chi_1^2$, with P values 0.383, 0.894, 0.397, 0.715, and 0.671 for $k = 2, 3,\ldots, 6$, showing no significant departures. These findings provide confirmation that the MTST asymptotic null distribution has one degree of freedom regardless of the number of traits included in the models.

## 4. Power analysis

### *4.1 Calculating the power using noncentrality parameter*

In assessing power and sample size requirements of tests, we calculated the noncentrality parameter (NCP) of the chi-square distribution of LRT under the alternative hypothesis, extending the methods used by Sham et al. (2000) for calculating the power of univariate tests. We first calculated $\lambda_T$, the total NCP required for achieving 80% power for $\alpha = 0.01$. For example, for MTST the value $\lambda_T = 11.6789$ can be calculated by solving the equation $1 - pchisq(6.635, ncp = \lambda_T, df = 1) = 0.8$, where 6.635 is the critical value for $\chi_1^2$ for $\alpha = 0.01$ and *pchisq* represents the chi-square cumulative distribution function. Then we need to find $\lambda^*$, the NCP value per family given the alternative hypothesis, from which we can obtain the required sample size $N^*$ by dividing $\lambda_T$ by $\lambda^*$ and can get the power by computing the value for $1 - pchisq(6.635, ncp = n \cdot \lambda^*, df = 1)$.

To explain how $\lambda^*$ is calculated, let $l(\theta) = -\frac{1}{2}\ln|\Sigma| - \frac{1}{2}(y-\mu)^T \Sigma^{-1}(y-\mu)$ denote the log likelihood and define $f(\theta) \equiv 2E_{\theta_1}[l(\theta)] = -\ln|\Sigma| - Tr(\Sigma^{-1}\Sigma_1)$ to be twice the expected log likelihood under the alternative $\theta_1$, where $\Sigma$ is a function of $\theta$, and $\Sigma_1$ represents the expected covariance matrix under the alternative hypothesis, obtained by plugging the true parameters $\theta_1$ into $\Sigma$. Then we have $\lambda^* = \sup_{\theta \in \Theta_1} f(\theta) - \sup_{\theta \in \Theta_0} f(\theta)$. Here, since the true value $\theta_1$ lies in the alternative hypothesis $\Theta_1$, the first term $\sup_{\theta \in \Theta_1} f(\theta)$ is simply $f(\theta_1)$. The second term $\sup_{\theta \in \Theta_0} f(\theta)$ is twice the maximum of the expected log-likelihood under the null hypothesis, , giving $\lambda^* = f(\theta_1) - f(\theta_0^*)$, where $\theta_0^*$ denote the maximizer of $f(\theta)$ over $\theta \in \Theta_0$. In obtaining $\theta_0^*$, we used the numeric optimization function "optim" in the R software.

The NCPs under both complete pleiotropic and general models were obtained in this way. Compared to the complete pleiotropic model, the only difference in treating the general model is that additional covariance parameters from the major gene effect component $A = (a_{ij})$ are also optimized in obtaining $\theta_0^*$, because of the increased dimension of the parameter $\theta$. For a type of model misspecification where the true model is a general model but a complete pleiotropic model was applied for conducting LRT, the NCP is calculated similarly, except two optimizations are required rather than one. To see the reason for this, letting $\theta_1$ lie in the general model and continuing to define $f(\theta) = 2E_{\theta_1}[l(\theta)]$, we have $\lambda^* = f(\theta_1^*) - f(\theta_0^*)$ where $\theta_0^*$ is defined as above but $\theta_1^* = \arg\sup_{\theta \in \Theta_1} f(\theta)$ is not simply $\theta_1$ in this case, because now $\theta_1$ lies in the general model, not in $\Theta_1$, which is the alternative hypothesis in the complete pleiotropic model.

### *4.2 Comparisons across different genetic models*

We needed to choose sets of parameter values in the alternative hypothesis at which to evaluate powers of the tests under consideration. We chose several sets of parameters as follows. We first defined three classes of models called A, B, and C with $k = 3$, varying the pattern of major gene effect size on each trait in the multivariate model. The model A assigned 40%, 5%, and 7% major gene effects for traits T1, T2, and T3, respectively, so that only one trait has a large major gene effect and the rest have relatively small effects. Model A was designed to mimic the pattern suggested by the results of Marlow shown in the first column of Table 1. In the model B, all traits had large major gene effect sizes (42%, 38%, and 40%), whereas in the model C, all traits had small major gene effects (7%, 5%, and 6%). For polygenic effects, 28%, 30%, and 32% were assigned traits T1, T2, and T3 in all of the models, and in each model the environmental effect was the remainder to total 100%.

Having chosen parameters describing the marginal (univariate) distributions of T1, T2, and T3 in each model class A, B, and C, the specification of our multivariate models were completed by choosing correlation coefficients in the matrices $G$ and $E$. We wanted to select several sets of correlation parameters that are well-spread out over the parameter space, and used the same method described by Han and Chang (2008). Using Inverse-Wishart distributions, we sampled 100 random pairs of covariance matrices $G$ and $E$ having uniformly distributed correlation coefficients, and then applied K-means clustering with three centers to select three sets of correlations. Combining each of these three sets of correlations (which we refer to as (a), (b), and (c)) with each of model classes A, B, and C produced nine multivariate models we will refer to as A-(a), A-(b), A-(c), B-(a), B-(b), …, C-(c).

We calculated powers and required sample sizes of the MTST and the UT on the nine genetic models described in the above. The results in Table 2 show there are overall power increases using the MTST relative to the UT across different models, with relative efficiencies mostly larger than one. For example, in model class A, achieving 80% power (with $\alpha = 0.01$) for the second trait T2 requires $N = 28956$ sib-pairs using the UT, compared to only $N = 1465$, 1775, and 1275 using the MTST in models A-(a), A-(b) and A-(c), respectively (so that the sample sizes are reduced by 19.76, 16.32 and 22.70). The corresponding powers calculated for n=1500 show the power for the UT increased from 0.06 to 0.81, 0.71 and 0.87 for the MTST in models (a), (b) and (c). We see the similar patterns in other traits and in the other model classes B and C.

Table 2. Powers (for sample size 1500) and required sample sizes (for 80% power) for the UT and the MTST across different genetic models. Each class of models A, B and C has three traits T1, T2, T3 with major gene effect sizes specified in the second column. Each model class contains three models ((a), (b), and (c)) having different sets of correlations among traits. For each trait and each model, we calculated required sample size ($N$) for 80% power with $\alpha = 0.01$, and power for sample size 1500 sib-pairs for the UT and the MTST. Relative efficiency (Effic) was defined to be the required sample size for the UT divided by the required sample size for the MTST.

| Model class | % | Trait | UT | | MTST | | | | | | | | |
|---|---|---|---|---|---|---|---|---|---|---|---|---|---|
| | | | | | Model (a) | | | Mode (b) | | | Mode (c) | | |
| | | | N | Power | N | Power | Effic | N | Power | Effic | N | Power | Effic |
| A | 40 | T1 | 330 | 1.00 | 250 | 1.00 | **1.32** | 346 | 1.00 | **0.95** | 203 | 1.00 | **1.63** |
| | 5 | T2 | 28956 | 0.06 | 1465 | 0.81 | **19.76** | 1775 | 0.71 | **16.32** | 1275 | 0.87 | **22.70** |
| | 7 | T3 | 14928 | 0.09 | 1060 | 0.93 | **14.08** | 1307 | 0.86 | **11.42** | 932 | 0.96 | **16.01** |
| B | 42 | T1 | 291 | 1.00 | 102 | 1.00 | **2.85** | 140 | 1.00 | **2.08** | 96 | 1.00 | **3.04** |
| | 38 | T2 | 367 | 1.00 | 110 | 1.00 | **3.35** | 148 | 1.00 | **2.48** | 104 | 1.00 | **3.54** |
| | 40 | T3 | 338 | 1.00 | 106 | 1.00 | **3.19** | 145 | 1.00 | **2.33** | 101 | 1.00 | **3.34** |
| C | 7 | T1 | 14765 | 0.09 | 3351 | 0.39 | **4.41** | 5859 | 0.20 | **2.52** | 2981 | 0.44 | **4.95** |
| | 5 | T2 | 28956 | 0.06 | 4454 | 0.28 | **6.50** | 7482 | 0.15 | **3.87** | 3996 | 0.31 | **7.25** |
| | 6 | T3 | 20432 | 0.07 | 3818 | 0.33 | **5.35** | 6555 | 0.17 | **3.12** | 3448 | 0.37 | **5.93** |

The patterns in Table 2 reveal that power gains for T2 or T3 are much larger than for T1 in model class A, where T2 and T3 have relatively small major gene effect sizes (5%, 7%) while T1 has the largest major gene effect (40%). T1 already has large power using the UT, and using the extra information in T2 and T3 through multivariate models did not seem be very helpful in increasing power. The analogous patterns are shown in model classes B and C, although the overall power gains are lower than in model class A. Note that model class A includes traits of substantially different major gene effect sizes, while model classes B and C include traits of similar effect sizes. Although all traits in model class C have small major gene effects like T2 and T3 in model class A, the power increases using the MTST in model class C were not as large as in model class A. Note that unlike model class A, model class C does not have a trait with substantially larger major gene effect than the other traits. From this, we hypothesized that including a trait with substantially larger major gene effect size helps increase the power of traits with smaller effect sizes. To investigate this idea, in the next section we study the impact of the major gene effect size for T1 on the power of the MTST for other traits (e.g., T2).

### *4.3 Varying the major gene effect size of another trait*

Focusing on the bivariate version of the model A-(a) in Table 2, which includes only T1 and T2, we varied the major gene effect size of T1 and calculated powers of the MTST for T2 to see the impact of another trait included in the multivariate model on the MTST. Figure 1 (a) shows the results in power (n=1500) and we see, by increasing the major gene effect size of T1 from 5% to 50%, the power for detecting the signal for T2 using the MTST increased from 0.15 to 0.90, and corresponding required sample sizes were reduced from 7260 to 1150 (red curves). On the other hand, the power of the UT (blue) stays constant since it is not using information from other traits in the model. This implies that when applying the MTST to a given trait, simply increasing the effect size of another trait included in the multivariate model can dramatically increase the power for the test, even if the effect size of the given trait remains the same. This is due to the fact that the NCP of the MTST for T2 is an increasing function of the major gene effect size of T1, as shown in Figure 1(c).

Figure 1. Impact of the major gene effect of T1 on the power of the MTST for T2. Red curves show results for the MTST and blue curves, for the UT. (a) Power for $N=1500$, (b) Required sample size for 80% power for $\alpha=0.01$, (c) NCP.

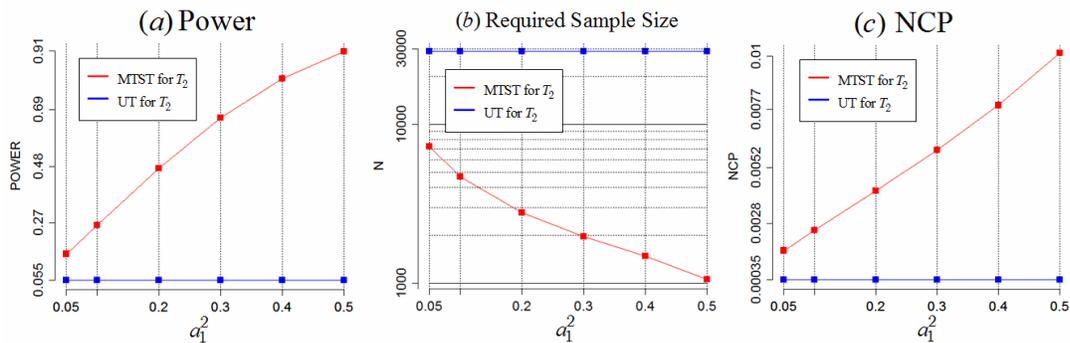

We then investigated how the impact of T1 on the MTST is influenced by other factors, focusing on the correlation between traits. In order to see if the impact of the T1 still remains without traits being correlated, we assigned zero correlation between two traits T1 and T2 in both the polygenic and environmental components, i.e. $\rho_{g_{12}} = \rho_{e_{12}} = 0$, in the model used in Figure 1 (that model previously had $\rho_{g_{12}} = 0.068$ and $\rho_{e_{12}} = -0.479$), and then repeated the calculation to get the NCP plot shown in Figure 1(c). The result in Figure 2(a) shows that even when T1 and T2 have zero correlation in G and E, the impact of the major gene effect of T1 does not disappear. A similar pattern as Figure 1(c) was observed, where the NCP of the MTST for T2 still increases with the major gene effect of T1 as before, although the overall NCP level is a bit lower than the original curve.

Figure 2. The impact of correlation in polygenic, environmental and major gene effects ($\rho_{g_{12}}, \rho_{e_{12}}$, and $\rho_{a_{12}}$). In plot (a), the results in Figure 1(c) were recalculated with zero correlation in G and E between traits. In plot (b), Figure 1(c) was recalculated under general models departing from complete pleiotropy, varying the major gene correlation from 0 to 0.9, with the red curve showing the earlier complete pleiotropy result.

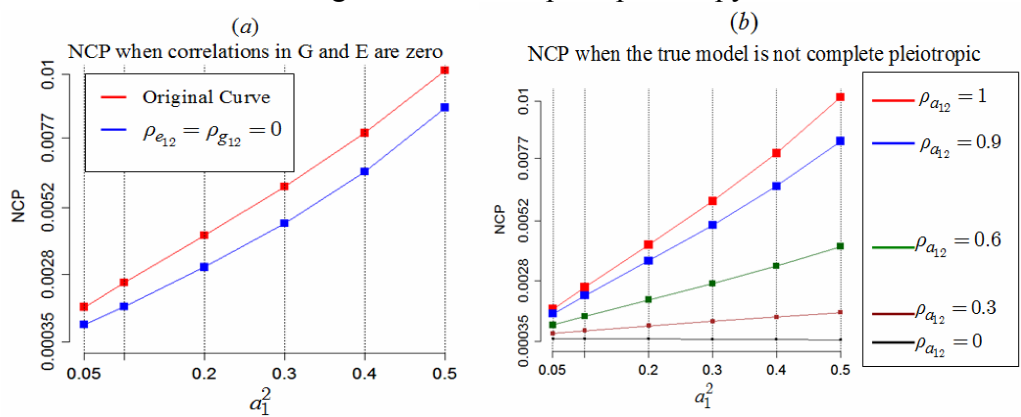

Note that even after setting $\rho_{g_{12}} = \rho_{e_{12}} = 0$, there still remains another source of correlation between two traits through the major gene effect component A, which, under the complete pleiotropy assumption, is perfectly correlated, that is, $\rho_{a_{12}} = \pm 1$. To explore how the curve in Figure 1(c) changes when the true model departs from the complete pleiotropic assumption, we considered a set of general models with $\rho_{a_{12}}$ taking values 0, 0.3, 0.6 and 0.9 and recalculated NCP for T2. The result in Figure 2(b) indicates, as the departure from complete pleiotropy grows (that is, as $\rho_{a_{12}}$ decreases), the contribution of the major gene effect of T1 on the power for T2 is reduced, and when $\rho_{a_{12}} = 0$, increasing the major gene effect size of T1 doesn't contribute to the power for T1 at all. The implication is that the power increase using the MTST is mainly due to the major gene correlation, and the test works more effectively when traits are highly correlated through the major gene effect.

### *4.4 Complete pleiotropy model vs. general model*

In the previous section we studied the case when the true model departs from complete pleiotropy, focusing on comparing NCP values for varying sizes of the major gene correlation. In calculating the NCP in the general models, we assumed the correct general model, which includes parameters for major gene correlation. However, since the MTST in current use in the literature assumes complete pleiotropy, we also wanted to examine the model misspecification effect. That is, we investigate the effect on power of applying an incorrectly specified complete pleiotropy model when the truth is a general model having imperfect major gene correlations, which is more realistic. One might expect a loss of power under this sort of model misspecification, but as we show below, in fact this is usually not the case.

Figure 3(a) considers a set of bivariate general models with $\rho_{a_{12}}$ taking values 0, 0.3, 0.5, 0.7 and 0.9, with the other parameters the same as in model A-(a). Required sample sizes are shown for the MTST for T2, when applying the general model (correct model specification) and when applying the complete pleiotropy model (incorrect model specification). When the major gene correlation is small (less than 0.35), that is, the model departs substantially from complete pleiotropy, the general model is more powerful and requires smaller sample sizes than the complete pleiotropy model. However, when the major gene correlation is larger, the complete pleiotropy model, is actually more powerful and requires smaller sample sizes than the general model, even though the complete pleiotropy model is misspecified. This can be understood because a general model has more parameters to be estimated, so that the asymptotic null distribution has increased degrees of freedom with increased critical values.

Figure 3. Comparing required sample sizes (for 80% power using $\alpha = 0.01$) of the MTST using a complete pleiotropy model (red) and a general model (blue) under the truth of general models with varying major gene correlations 0, 0.3, 0.5, 0.7 and 0.9. The MTST was conducted for T2 in a bivariate model in plot (a), and in a $k = 6$ trait multivariate model in plot (b).

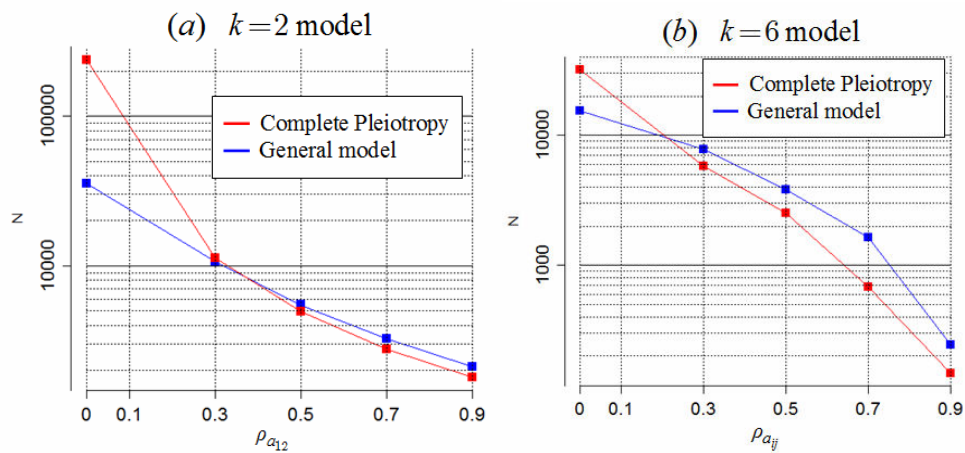

For the bivariate model, we found the asymptotic null distribution of the MTST for a general model is a mixture of $\chi_1^2$ and $\chi_2^2$ with equal probabilities, i.e. $\frac{1}{2}\chi_1^2 + \frac{1}{2}\chi_2^2$, rather than $\chi_1^2$, the asymptotic null distribution under complete pleiotropy. Thus the critical value used in Figure 3(a) for a general model is 8.278 for $\alpha = 0.01$, which is larger than the corresponding critical value 6.635 for a complete pleiotropy model. This mixture distribution for the general model was obtained using analogous geometric arguments as in Han and Chang (2008), and simulations were conducted to confirm this. The same simulation method was used as described in Section 2.2, but under a general model, and the simulated distribution was compared to $\frac{1}{2}\chi_1^2 + \frac{1}{2}\chi_2^2$. The Kolmogorov-Smirnov test gave P value 0.657, and estimated type 1 error rates for $\alpha = 0.01$ and 0.05 were 0.010, and 0.048, which are reassuring.

Increasing the number of traits in a multivariate model increases the number of covariance parameters to be estimated in a general model, hence raising the degrees of freedom of the asymptotic null distribution. Our simulation results show the critical values of the $k$-trait general model increase over $k$. For example, for $\alpha = 0.01$, the critical values were 8.912, 9.385, 10.827, and 11.818 for $k = 3, 4, 5, 6$ respectively.

In order to see how the size of a model influences the misspecification effect (that is, the relative performances of the general model and the complete pleiotropy model under the truth of a general model), we repeated the procedure done in Figure 3(a) using $k = 6$ trait models. Four more traits were added to the previous bivariate model, with major gene effect sizes 15%, 25%, 20%, and 30%, and polygenic effects equally assigned 30% for all four traits. A set of correlation coefficients to use in G and E was randomly sampled from an Inverse-Wishart distribution as in Section 2.4, and the major gene correlation among all traits took values 0, 0.3, 0.5, 0.7 and 0.9. The result in Figure 3(b) shows that using the complete pleiotropy model becomes more advantageous in a bigger model and shows more powerful performance than the general model at a lower level of major gene correlations than in Figure 3(a).

## 5. Application of a power calculator to reading disability data

We implement the MTST for testing shared gene effects in an R package "sharedGene" that includes a power calculator that can be used for designing studies by using pilot data to calculate required sample sizes and to choose traits for further studies. The package uses the software Mx for conducting LRT and estimating parameters, which are used for power calculations. We applied this R package to the data on reading disability in the Russian language.

The data were collected in a study conducted by the Yale Child Study Center in the town of Voronezh in Russia, by randomly selecting 15 elementary schools, and identifying students with reading difficulties who had at least one sibling attending the same school. This process yielded a total of 352 probands in 352 families, among which 336 families had 2 sibs, 15 had 3 sibs and 1 had 4 sibs. The phenotypes under study were five reading-related quantitative measures including phonological, orthographic, and morphological awareness in addition to spelling and comprehension skills. For genetic data, we analyzed 35 microsatellite markers on chromosome 3. IBD proportions were estimated using Merlin multipoint method (Abecasis et al., 2001).

We identified a locus on 3q29 (the nearest marker D3S1311), by conducting the UT across chromosome 3, where the most significant P value was 0.00001 for the morphology trait (see Table 3). This locus is near regions reported to influence written and spoken language disorders in other languages including Finnish (Nopola-Hemmi et al., 2001) and English (Stein et al., 2004). According to the UT results, other traits besides morphology did not show any significance (with P values between 0.14 and 0.5), suggesting the effect of this locus is specific to morphological skills. The results using the MTST, however, showed increased significance for the other traits as well, with P values decreasing to 0.00114 ~ 0.02674, suggesting the possibility that the 3q29 locus is a shared common genetic risk factor for all of our traits related to reading disability in the Russian language.

The corresponding power calculations for these tests explain this increased significance, as shown in Table 3. For example, the power of the spelling trait was raised from 0.23 in the UT to 0.80 in the MTST, which paved the way for the corresponding P values to reduce from 0.41578 to 0.00114. The major gene correlations between all traits were estimated to be quite high, ranging from 0.43 to 0.96 in absolute value, which is consistent with our results from Section 3.2 that the MTST shows increased power when traits are highly correlated through a major gene component.

Table 3. Application of the MTST, the UT, and the power calculator to reading disability data. The P values and the power were calculated for the 3q29 locus. Required sample size ($N$) was calculated for 80% power with $\alpha = 0.01$, and power was calculated for sample size 352 sib-pairs for the UT and the MTST. Relative efficiency (Effic) was defined to be the required sample size for the UT divided by the required sample size for the MTST.

| Trait | UT P value | MTST P value | UT $N$ | UT Power (n=352) | MTST $N$ | MTST Power (n=352) | Effic |
|---|---|---|---|---|---|---|---|
| Morphology | 0.00001 | 0.00042 | 246 | 0.920 | 236 | 0.938 | 1.04 |
| Orthography | 0.14425 | 0.02674 | 7379 | 0.051 | 605 | 0.497 | 12.19 |
| Phonology | 0.50000 | 0.02341 | 9908 | 0.042 | 746 | 0.396 | 13.28 |
| Spelling | 0.41578 | 0.00114 | 1360 | 0.230 | 335 | 0.809 | 4.06 |
| Comprehension | 0.33007 | 0.02000 | 5500 | 0.063 | 530 | 0.567 | 10.38 |

In fact, the sample of size $n = 352$ described above is pilot data, and we intend to design a further study by collecting more samples, using the power calculation results reported in Table 3. From the UT power results we see that achieving 80% power for all five traits with significance level $\alpha = 0.01$ requires $n = 9908$, so that an additional 9556 sib-pairs need to be collected. The power calculations based on the MTST, however, give $n = 746$, reducing the required sample size by a factor of more than 13. The power calculator can help in other ways in designing an efficient follow-up study. For example, suppose we would like to select a subset of, say, 3 out of the 5 traits, in such a way that the subset has small required sample size. To do this we can calculate powers for all possible models having 3 traits (10 models all together). Our results show the model including three traits morphology, orthography, and spelling gives the most efficient design with required sample size 494 for achieving 80% power for all three traits.

## 6. Discussion

The MTST introduced in Marlow's study has had substantial impact in the literature of neurodevelopmental, cognitive, and psychiatric disorders (Bates, 2006; Cherny, 2008; Consortium, 2004; Fisher, 2006; Hansell et al., 2005; Loo et al., 2004; Monaco, 2007; Schulte-Korne, 2002; Wright et al., 2008; Zhao et al., 2008) and has been noted to be the first multivariate method to provide clarification of the patterns of QTL influences on complex traits (Hansell et al., 2005). The impact of Marlow's study has extended beyond neurodevelopmental disorders to other complex disorders in which studying shared gene effects on multiple aspects of the disorder is of great interest, including allergic disorders such as asthma (Bouzigon et al., 2004; Bouzigon et al., 2007), eczema (Guilloud-Bataille et al., 2008), obesity (Zhao et al., 2008), and platelet disorder (Evans et al., 2004).

This paper has provided the first study of the validity and power of the MTST. We have shown the MTST using the chi-square with one degree of freedom as the null distribution has correct type 1 error rates regardless of the number of traits in the model. Our NCP calculations also established power increases compared to standard univariate methods across different genetic models. In addition, we find an explanation for the main source of these power gains: they arise due to correlations among traits induced by a common major gene effect component. We also compared the use of the complete pleiotropy model, as assumed by Marlow, to the general model allowing additional correlation parameters. Previous work (Amos et al., 2001) has expressed concerns about using the complete pleiotropy model, although this work considered another type of multivariate test that jointly tests the parameters for all traits. However, the conclusions of this work were unclear, employing, in addition to simulations, binomial mixing probabilities in null distributions both for the complete pleiotropy and general models, which have been shown to be incorrect (Han and Chang, 2008). Here, using clarified null distributions both for the complete pleiotropy and general models, we find that using the complete pleiotropy model can be more powerful even in certain model misspecification situations in which the true model is a general model, as long as traits are moderately correlated by a major gene component.

Our power calculator, implemented in freely available software, was also shown to be helpful in designing efficient studies, calculating required sample sizes, and choosing traits for further studies. Power calculations based on standard univariate tests do not fully use information from correlations among traits, and so when applied to our data, showed required sample sizes that were more than 13 times larger than those for our method based on the MTST. We also demonstrated how to improve efficiency by calculating required sample sizes for subsets of traits and choosing the subset having the smallest sample size.

Studies aiming to uncover genetic overlaps among multiple aspects of disorders have not been limited to studying a single disorder at a time. Cross-disorders approaches to identifying genetic overlaps have been applied to bipolar and schizophrenia (Craddock, O'Donovan and Owen, 2005; Murray et al., 2004), ADHD and reading disability (Loo et al., 2004; Stevenson et al., 1993), and among diverse cognitive skills including mathematics, languages, and general cognitive skill such as memory and spatial ability (Haworth et al., 2007; Kovas and Plomin, 2006, 2007; Plomin, Kovas and Haworth, 2007). Most of these studies, however, either compare linkage peaks in separate univariate scans (Craddock et al., 2005; Loo et al., 2004), or focus on polygenic models by estimating trait correlations across different disorders without using genetic marker information (Kovas and Plomin, 2007; Stevenson et al., 1993). We believe our study of the power and validity of the MTST and our implementation of these methods can be of great use for researchers seeking common shared genetic risk factors among aspects of complex disorders, including cross-disorder studies, both for conducting the tests and for designing efficient studies.

Nopola-Hemmi, J., Myllyluoma, B., Haltia, T*., et al.* (2001). A dominant gene for developmental dyslexia on chromosome 3. 658-664: BMJ.

Plomin, R., Kovas, Y., and Haworth, C. M. A. (2007). Generalist Genes: Genetic Links Between Brain, Mind, and Education. *Mind, Brain, and Education* **1**, 11-19.

Schmitz, S., Cherny, S. S., and Fulker, D. W. (1998). Increase in Power through Multivariate Analyses. *Behavior Genetics* **28**, 357-363.

Schulte-Korne, G. (2002). Annotation: Genetics of reading and spelling disorder. *The Journal of Child Psychology and Psychiatry and Allied Disciplines* **42**, 985-997.

Self, S. G., and Liang, K. Y. (1987). Asymptotic properties of maximum likelihood estimators and likelihood ratio tests under nonstandard conditions. *J Am Stat Assoc* **82**, 605-610.

Sham, P. C., Cherny, S. S., Purcell, S., and Hewitt, J. K. (2000). Power of Linkage versus Association Analysis of Quantitative Traits, by Use of Variance-Components Models, for Sibship Data. *The American Journal of Human Genetics* **66**, 1616-1630.

Smalley, S. L. (1997). Genetic influences in childhood-onset psychiatric disorders: autism and attention-deficit/hyperactivity disorder. *The American Journal of Human Genetics* **60**, 1276-1282.

Stein, C. M., Schick, J. H., Gerry Taylor, H*., et al.* (2004). Pleiotropic Effects of a Chromosome 3 Locus on Speech-Sound Disorder and Reading. *The American Journal of Human Genetics* **74**, 283-297.

Stevenson, J., Pennington, B. F., Gilger, J. W., DeFries, J. C., and Gillis, J. J. (1993). Hyperactivity and Spelling Disability: Testing for Shared Genetic Aetiology. *Journal of Child Psychology and Psychiatry* **34**, 1137-1152.

Vogler, G. P., Tang, W., Nelson, T. L*., et al.* (1997). A multivariate model for the analysis of sibship covariance structure using marker information and multiple quantitative traits. *Genetic Epidemiology* **14**, 921-926.

Williams, J. T., Begleiter, H., Porjesz, B*., et al.* (1999a). Joint Multipoint Linkage Analysis of Multivariate Qualitative and Quantitative Traits. II. Alcoholism and Event-Related Potentials. *The American Journal of Human Genetics* **65**, 1148-1160.

Williams, J. T., Van Eerdewegh, P., Almasy, L., and Blangero, J. (1999b). Joint Multipoint Linkage Analysis of Multivariate Qualitative and Quantitative Traits. I. Likelihood Formulation and Simulation Results. *The American Journal of Human Genetics* **65**, 1134-1147.

Wright, M. J., Luciano, M., Hansell, N. K., Montgomery, G. W., Geffen, G. M., and Martin, N. G. (2008). QTLs Identified for P3 Amplitude in a Non-Clinical Sample: Importance of Neurodevelopmental and Neurotransmitter Genes. *Biological Psychiatry* **63**, 864-873.

Zhao, J., Xiao, P., Guo, Y. A. N*., et al.* (2008). Bivariate genome linkage analysis suggests pleiotropic effects on chromosomes 20p and 3p for body fat mass and lean mass. *Genetics Research* **90**, 259-268.
17